\documentclass[twocolumn,aps,prd,showpacs,amsmath,amssymb,superscriptaddress,nofootinbib]
{revtex4}
\usepackage{amsmath,amsfonts,amssymb}
\usepackage{latexsym}
\usepackage{dcolumn}
\usepackage{graphicx,epsfig}
\usepackage{amsthm}
\usepackage{color}
\usepackage{hyperref}

\begin{document}

\title{Emergent universe in Ho\v{r}ava-Lifshitz-like $F(R)$ gravity}

\author{M. Khodadi}
\email{m.khodadi@iaufb.ac.ir}
\affiliation{ Young Researchers and Elite Club, Firoozkooh Branch, Islamic Azad
University,
Firoozkooh, Iran}

\author{ Y. Heydarzade}
\email{heydarzade@azaruniv.edu} \affiliation{Department of Physics, Azarbaijan Shahid
Madani University, Tabriz,
53714-161,
Iran}

\author{ F. Darabi}
\email{f.darabi@azaruniv.edu} \affiliation{Department of Physics, Azarbaijan Shahid
Madani University, Tabriz,
53714-161,
Iran}
\affiliation{ Research Institute for Astronomy and Astrophysics of Maragha (RIAAM),
Maragha
55134-441, Iran}

\author{E. N. Saridakis}
\email{Emmanuel\_Saridakis@baylor.edu}
\affiliation{CASPER, Physics Department, Baylor University, Waco, TX 76798-7310, USA}
\affiliation{Instituto de F\'{\i}sica, Pontificia
Universidad de Cat\'olica de Valpara\'{\i}so, Casilla 4950,
Valpara\'{\i}so, Chile}

\begin{abstract}
We investigate the Einstein static universe (ESU) and the emergent universe scenario in 
the framework of Ho\v{r}ava-Lifshitz-like $F(R)$ gravity. We first perform a dynamical 
analysis in the phase space, and amongst others we show that a spatially open universe 
filled with matter satisfying the strong energy condition can exhibit a stable static 
phase. Additionally, we examine the behavior of the scenario under scalar perturbations
and extract the conditions under which it is free of perturbative instabilities,  showing 
that the obtained background ESU solutions are free of such instabilities. However, 
in order for the Einstein static universe to give rise to the emergent universe scenario  
we need to have an exotic matter sector that can lead the universe to depart from the 
stable static state and enter into its usual expanding thermal history.
\end{abstract}

\maketitle

%%%%%%%%%%%%%%%%%%%%%%%%%%%%%%%%%%%%%%%%%%%%%%%%%%%%%%%%%%%%%%%%%%%
\section{Introduction}

According to the scenario of old Big-Bang cosmology based on the theory of general
relativity (GR), our universe has begun from a finite past including an initial
singularity, which is widely considered as a conceptual disadvantage. Furthermore, since
standard Big-Bang cosmology was incapable of solving the horizon, flatness and magnetic
monopole problems at early universe the inflation mechanism was introduced
\cite{Inflation}. Finally, in order to describe the late-time acceleration a small
positive cosmological constant was added, giving rise to the Standard Model of the
Universe, namely $\Lambda$CDM cosmology. However, despite the remarkable achievements of
this paradigm, its physical content, concerning both early and late-time accelerating
phases, is still ambiguous, and moreover amongst others one still faces the ``initial
singularity problem''.

Concerning the initial singularity problem, one could try to confront it through  the
``emergent universe'' scenario \cite{Ellis}. In particular, in such a scenario the
universe is originated from a static state, known as ``Einstein static universe'' (ESU),
and then it entered the inflationary phase, without ever passing through the Big-Bang
singularity. However, remaining in the GR framework, and despite fine-tuning, perturbation
analysis shows that the initial singularity cannot be completely removed.
Indeed, ESU is severely influenced by the initial conditions such as perturbations which
are prevailed in the Ultra-Violet (UV) limit, and it is indeed unstable against classical
perturbations which eventually make it collapse towards a singularity \cite{Eddington}.

In order to alleviate the above problems, alternative cosmological models have been
developed. A first direction is to introduce new, exotic, forms of matter, which in the
framework of general relativity could provide an explanation of the observed universe
behavior \cite{Copeland:2006wr,Cai:2009zp}. A second direction is to modify the
gravitational sector, obtaining a theory that still possess GR as a particular limit,
but still being able to describe the universe at large scales through the extra
gravitational degrees of freedom \cite{Nojiri:2006ri,Capozziello:2011et}. Indeed, in
modified gravity, amongst others one can cure the emergent universe scenario, by making
stable
Einstein static universe. Hence, a large amount of activity was devoted to the
study of the stability of the Einstein static universe in various gravitational
modifications, such as Einstein-Cartan theory \cite{EC}, Lyra geometry \cite{sadra},
non-constant pressure models \cite{ncp}, $f(R)$ gravity \cite{f(R)}, $f(T)$ gravity
\cite{f(T)}, loop quantum cosmology \cite{LQC}, massive gravity \cite{MG} and doubly
general relativity \cite{DGR}, induced matter theory \cite{IMT}, braneworld
models \cite{brane}, etc.

One interesting gravitational modification, proposed by  Ho\v{r}ava, is the so-called
Ho\v{r}ava-Lifshitz gravity \cite{Peter}. This construction was motivated by the
observation that the insertion of higher-order derivative terms in the Einstein-Hilbert
Lagrangian establishes renormalizability, since the graviton propagator at high energies
is modified \cite{Stelle:1976gc,Biswas:2011ar}. Nevertheless, this leads to a severe
problem, since the equations of motion involve higher-order time derivatives and hence the
theory includes ghosts. However, since it is the higher spatial derivatives that
improve renormalizability while  it is the higher time derivatives that lead to ghosts,
one could think of constructing a theory that would allow for the inclusion of higher
spatial derivatives only. Indeed, this is what it is achieved in Ho\v{r}ava-Lifshitz
theory \cite{Peter}, and since higher spatial derivatives are
not accompanied by higher time ones, in the UV the theory exhibits power-counting
renormalizability but still without ghosts. Such a theory definitely violates Lorentz
invariance, however it presents GR as an Infra-Red fixed point, where Lorentz invariance
is restored. Application of Ho\v{r}ava-Lifshitz gravity in a cosmological framework leads
to very interesting behavior in agreement with observations \cite{Review}. Finally, one
can proceed further by construction extensions such as $F(R)$ Ho\v{r}ava-Lifshitz gravity
\cite{Kluson1}, since in such scenarios one can obtain a unified mechanism for the
early-time inflation and the late-time acceleration \cite{Chaichian1}.

From the above discussion we deduce that it is worthy to examine the realization of ESU
and of the emergent universe scenario in the framework of  Ho\v{r}ava-Lifshitz gravity
and its extensions. The stability issues of the ESU  in the framework of an IR
modification of Ho\v{r}ava-Lifshitz gravity, representing a soft breaking of the
so-called ``detailed balance condition'', against linear homogeneous scalar
perturbations, was explored in \cite{HL}. As it is shown, there exists a large class of
stable solutions, for large regions of barotropic equation-of-state
parameter and model parameters, however the possibility for a transition to the
inflationary era is ambiguous. Additionally, in the
context of original Ho\v{r}ava-Lifshitz gravity such a study was performed in \cite{P},
where it was shown that a stable ESU can be realized in the presence of a negative
cosmological constant, however although  the Big Bang singularity can be avoided the
transition from this stable state to the inflationary era is impossible. On the other
hand, in the case of the generalized version of Ho\v{r}ava-Lifshitz gravity \cite{T}, it
was shown that if the cosmic scale factor satisfies certain conditions initially and if
the equation-of-state parameter approaches a critical value, the corresponding
stable critical point coincides with the unstable one, and consequently a phase transition
to the inflationary era can be provided \cite{P}.

In the present work we are interested in studying the emergent universe scenario in the
framework of a Ho\v{r}ava-Lifshitz-like $F(R)$ gravity, and in particular in
investigating the realization and stability of the Einstein static universe
and the possibility of the phase transition to the inflationary era. The plan of the
manuscript is the following: In section \ref{themodel}, we briefly review
Ho\v{r}ava-Lifshitz-like $F(R)$ gravity and we apply it in a
cosmological framework. In section \ref{phasespaceanalysis}, we perform a dynamical
analysis in the phase space, while in section \ref{perturbationsanalysis}, we extract the
conditions under which the scenario at hand is stable against scalar perturbations.
Finally, section \ref{Conclusions} is devoted to discussion and conclusions.

\section{Ho\v{r}ava-Lifshitz-like $F(R)$ gravity and cosmology}
\label{themodel}

In this section we review the $F(R)$ Ho\v{r}ava-Lifshitz theory \cite{Kluson1}
(see also \cite{Chaichian1, Kluson2, Gourgoulhon}). In
this construction one starts from the usual Ho\v{r}ava-Lifshitz gravity \cite{Peter}  and
adds the $F(R)$ sector \cite{FR}, namely one replaces the Ricci scalar $R$ by arbitrary
functions of it.

We start by using the Arnowitt-Deser-Misner (ADM) formalism in a (3+1)
space-time \cite{ADM}, writing the metric as
\begin{equation}
\label{1,1}
ds^{2}=-N^{2}dt^{2}+h_{ij}(dx^{i}+N^{i}dt)(dx^{j}+N^
{j}dt),
\end{equation}
with $i,j=1,2,3$, where $N$, $N^{i}$ and $h_{ij}$ are respectively the lapse
function, the shift function and the metric of three-dimensional spatial hypersurface
$\Sigma_{t}$. In the framework of standard $F(R)$ gravity, the modified action is
\cite{Chaichian1}
\begin{equation}
\label{2,1}
S_{F(R)}=\int d^{4}x\,\sqrt{-g}\,\,F(R),
\end{equation}
with $\sqrt{-g}=\sqrt{h}N$, where $ F(R)$ denotes an arbitrary function of the
scalar curvature $R$. This is given by
\begin{equation}\label{3,1}
R=R^{(3)}+K^{ij}K_{ij}-K^{2}+2\nabla_{\mu}(n^{\mu}\nabla_{\nu}n^{\nu}-n^{\nu}
\nabla_{\nu}n^{\mu})~,
\end{equation}
where $R^{(3)}$, $K_{ij}$ and $n^{\mu}$ are the three-dimensional scalar
curvature, extrinsic curvature and a unit vector perpendicular to the three-dimensional
spacelike hypersurface $\Sigma_{t}$, respectively. The
term $R^{(3)}$ is an object associated with the spatial metric $h_{ij}$ of
the hypersurface, and $K_{ij}$ is defined as
\begin{equation}\label{4,1}
K_{ij}=\frac{1}{2N}(\dot{h_{ij}}-\nabla_{i}^{(3)}N_{j}-\nabla_{j}^{(3)}N_{i})~.
\end{equation}
Hence, one can now write down the action of $F(R)$ Ho\v{r}ava-Lifshitz gravity as
\cite{Chaichian1, Kluson2}
\begin{equation}\label{5,1}
S_{F(\tilde{R})}=\int dtd^{3}x\sqrt{h}\,N\,F(\tilde{R})~,
\end{equation}
with the extended scalar curvature $\tilde{R}$ given by
\begin{eqnarray}
\label{6,1}
&&\tilde{R}\equiv K^{ij}K_{ij}-\lambda K^{2}+2\mu\nabla_{\mu}(n^{\mu}\nabla_{\nu}n^
{\nu}-n^{\nu}\nabla_{\nu}n^{\mu})\nonumber\\
&&\ \ \ \ \ \ -E^{ij}{\cal{G}}_{ijkl}E^{kl}~.
\end{eqnarray}
In the above expression the running dimensionless constant $\lambda$ appears due to the
``super-metric'', defined on the hypersurface $\Sigma_{t}$ as
\begin{equation}
\label{7,1}
{\cal{G}}^{ijkl}=\frac{1}{2}\left(h^{ik}h^{jl}+h^{il}h^{jk}\right)-\lambda h^{ij}
h^{kl}~,
\end{equation}
while $E^{kl}$ has been inserted in order to embed the satisfaction of the ``detailed
balance condition'', which is defined by an action $W[h_{kl}]$ on $\Sigma_{t}$ as
\begin{equation}
\label{8,1}
\frac{\delta W[h_{kl}]}{\delta h_{ij}}=\sqrt{h}E^{ij}\;.
\end{equation}
The detailed balance condition was inspired by the condensed matter physics \cite{cond}
and it implies that the potential term in the Lagrangian of a $D + 1$ dimensional theory
is derivable from the variation of the $D$ dimensional action \cite{Peter}. In
particular, while the shift variables $N^{i}(t,\textbf{x})$ and spatial metric
$h_{ij}(t,\textbf{x})$  are function of both space and time,  the lapse variable is
assumed to be a function of time only, namely $N=N(t)$, an assumption compatible
with the foliation preserving diffeomorphism. \\

In the initial formulation of $F(R)$ Ho\v{r}ava-Lifshitz gravity \cite{Kluson1, Kluson2}
Kluson focused on a Lagrangian of a new class of $F(R)$ gravity theories with
respect to the projectability condition, namely
\begin{equation}
\label{8,*}
{\cal{L}}=N\sqrt{g}\bigg[\frac{2K_{ij}{\cal{G}}^{ijkl}K_{kl}}{F^{\,'}(A)\kappa^{2}}
-\kappa^{2}F(A)\bigg],
\end{equation}
where $F(A)$ is an arbitrary function of the auxiliary field $A$, which   is a
function
of $K_{
ij}$ and $g_{ij}$, namely
\begin{equation}
\label{8,**}
A=\bigg(\frac{E^{ij}{\cal{G}}_{ijkl}E^{kl}+4\kappa^{-4}K_{ij}{\cal{G}}^{ijkl}K_{kl}}
{1-4\kappa^{-4}K_{ij}{\cal{G}}^{ijkl}K_{kl}}\bigg).
\end{equation}
By setting the typical function $F(A)=\sqrt{1+A}-1$, the corresponding action
$S=\int dt d^{3}x {\cal{L}}$
reads as
\begin{eqnarray}
\label{9,1}
&&\!\!\!\!\!\!\!\!\!\!\!\!\!\!\!\!
S=-\kappa^{2}\int dt d^{3}x\sqrt{h}N\left\{
\sqrt{1+\frac{1}{4}E^{ij}{\cal{G}}_{ijkl}E^{kl}}\right.\nonumber\\
&&\ \
\left.\cdot
\sqrt{1-\frac{4}{\kappa^{4}}(K^{ij}K_{ij}-\lambda K^{2})}-1\right\}
+S_m,
\end{eqnarray}
with  $\kappa^{2}\equiv32G\pi c$    the gravitational constant, set for simplicity to $1$
in
the following.
%Note that above action written in terms of natural unite which otherwise
%is necessary to $M^6$ ($M$ is mass parameter) be replaced rather than unite.
In the above action we have also considered the matter sector
characterized by $S_m$. Additionally, one could extend this action in a
non-projectable version \cite{Kluson2}. We mention that the scenario of  action
(\ref{9,1}) is invariant under the foliation preserving diffeomorphism and not under the
full  $D + 1$ diffeomorphism, which is in contrast to the usual $F(R)$ gravity theories.

The linearized version of the action (\ref{9,1}) writes as
\begin{eqnarray}
\label{9,1*}
&&\!\!\!\!\!\!\!\!\!\!\!\!\!\!\!\!
S=\int dt d^{3}x N\sqrt{g}\left[\frac{2}{\kappa^{2}}(K_{ij}K^{ij}-\lambda
K^{2})\right.\nonumber\\
&& \ \ \ \ \ \ \ \ \ \ \ \ \left.-
\frac{\kappa^{2}}{8}E^{ij}{\cal{G}}_{ijkl}E^{kl}\right],
\end{eqnarray}
which as expected reproduces the original Ho\v{r}ava-Lifshitz action. On the other hand,
note that by choosing  $\lambda=\mu=1$ in   (\ref{6,1}),
one obtains the usual $F(R)$ gravity. Therefore, the non-linear
action (\ref{9,1}) is a form of   Ho\v{r}ava-Lifshitz-like $F(R)$
gravity (see  \cite{Kluson1, Chaichian1, Kluson2}). More precisely, the Lagrangian
density of the action (\ref{9,1}) is not of the exact $F(R)$ form with any modified
scalar curvature expression. Instead, it belongs to a class of models where the
Lagrangian depends on the kinetic term $K_{ij}
\mathcal{G}^{ijkl}K_{kl}$ and the potential term $E^{ij}\mathcal{G}_{ijkl}E^{kl}$ in a
different way, and not only on the sum of the kinetic   and the potential terms, as in
the modified scalar curvature (\ref{6,1}) (note that such more general modified
Ho\v{r}ava-Lifshitz theories, where the
Lagrangian  can depend on the kinetic and potential terms independently, have also
been considered in \cite{Chaichian1}).

In order to apply the above theory in a cosmological framework, that is in order to
construct Ho\v{r}ava-Lifshitz-like $F(R)$ cosmology, we impose an FRW metric
on the 3-hypersurface $\Sigma_t$, namely
 \begin{equation}
 \label{10-1}
h_{ij}dx^{i}dx^{j}=a(t)^{2}\left[\frac{dr^{2}}{1-kr^{2}}+r^{2}(d\theta^{2}+\sin^{2}
\theta d\varphi^{2})\right]~,
\end{equation}
where $a(t)$ is the scale factor and with $k=-1,\,0$ and $1$ corresponding to open, flat
and closed universe, respectively. Furthermore, concerning the matter content
of the universe, we consider it to be a perfect fluid with energy-momentum tensor in
co-moving coordinates given by
\begin{equation}
\label{T}
 T_{\mu\nu}=(\rho + p)u_{\mu}u_{\nu}+ pg_{\mu\nu},
\end{equation}
where $u_{\alpha}=\delta^{0}_{\alpha}$ is the four-velocity vector of
the fluid, and $\rho$ and $p$ are the energy density and isotropic
pressure, respectively. Under these assumptions, the field equations derived from
action (\ref{9,1}) give rise to the Friedmann equations, which write as
\cite{Sayan}:
{\small{
\begin{equation}
\label{1,a}
\!\!\!
H^{2}=\frac{1}{6(3\lambda-1)(1-\rho)^2}\left[\rho-\frac{1}{2}\rho^{2}+6\Lambda_{w}-
\frac{12k}{a^{2}}+\frac{6k^{2}}{\Lambda_{w}a^{4}}\right],
\end{equation}
 \begin{eqnarray}
 \label{2,a}
&&\!\!\!\!\!\!\!\!\!\!\!\!
\dot{H}=\frac{1}{4(3\lambda-1)(1-\rho)^2}\left[(\rho+p)(\rho-1)+\frac{8k}{a^{2
}}+\frac{8k^{2}}{|\Lambda_{w}|a^{4}}\right]\nonumber\\
&& \, -3(\rho+p)H^{2}
(1-\rho)^{-1}.
\end{eqnarray}}}
The parameter $\Lambda_{\omega}$ is the effective cosmological constant, which similarly
to the original version of Ho\v{r}ava-Lifshitz model must be negative for the running
coupling parameter
$\lambda>\frac{1}{3}$. We mention that the above equations include a negative energy
density squared term $\rho^2$, which can have a significant role for the early stage of
the universe. Technically speaking, the presence of $\rho^2$ term in the
above Friedmann
equations is a direct upshot of nonlinear dependence to matter source in
this modified gravity theory
in the early universe. A detailed discussion on this subject is presented in \cite{Sayan}.
Finally,
let us point out that the value of $\lambda$ is commonly divided into two ranges, namely
$0<\lambda<\frac{1}{3}$ and $\lambda>\frac{1}{3}$. However, phenomenological studies
analyzing the observational data suggest that the value of $\lambda$ is constrained
in a narrow range around $\lambda=1$, i.e. $|\lambda-1|\lesssim0.02$
\cite{Dutta:2009jn}, which was expected since GR is obtained for $\lambda=1$. Hence, in
the following we restrict our analysis in the regime $\lambda>\frac{1}{3}$.

The cosmological application of Ho\v{r}ava-Lifshitz-like $F(R)$ gravity proves to have
many interesting features \cite{Chaichian1}. Amongst others it can provide a unified
mechanism for the
description of both the early-time inflation and the late-time acceleration, or
alternatively it can give rise to bouncing solutions \cite{Sayan} that can cure the
initial-singularity problem. Hence, it would be interesting to study if one can naturally
obtain a stable Einstein static universe in such a theory. We mention that, as it is
well-known, the versions of  Ho\v{r}ava-Lifshitz gravity with the projectability condition
suffer from the strong-coupling problem \cite{Bogdanos:2009uj}. Although this can be
alleviated by going beyond the
projectability condition \cite{Blas:2009qj}, the corresponding $F(R)$ extension would be
too difficult to
allow for an analytical investigation of the emergent universe scenario. Hence, in the
present work
we consider the Ho\v{r}ava-Lifshitz-like $F(R)$ gravity with  the detailed balance and
projectability condition as a first approach on the background cosmological evolution,
having in mind that the investigation of the full theory is necessary as a next step.

\section{Dynamical analysis of Einstein static universe }
\label{phasespaceanalysis}

In this section we intend to perform a phase-space analysis, investigating non-flat
Ho\v{r}ava-Lifshitz-like $F(R)$
cosmology as a first-order cosmological dynamical system. This study will
show whether the Einstein static universe (ESU) corresponds to a solution in which the
universe can remain for very large time intervals. If this is not the case, then  the
realization of ESU will be highly improbable without a fine
tuning of the initial conditions.

Using Taylor expansion in terms of
$\frac{\rho}{\kappa^{2}}$ (note that we have set $\kappa^2=1$), the first and second
Friedmann equations (\ref{1,a}),(\ref{2,a}) can be rewritten as
\begin{eqnarray}
\label{1,b}
&&\!\!\!\!\!\!\!\!\!\!\!\!\!\!\!\!
H^{2}=\frac{1}{6(3\lambda-1)}\left[\rho-\frac{5}{2}\rho^{2}
+3\Lambda_{w}(\rho^2-\rho+1)\right.
\nonumber\\
&&\left.  \!\!\!
+ 6\left(\frac{k^{2}}{\Lambda_{w}a^{4}}-\frac{2k}{a^{2}}\right)
 (3 \rho
^2-2\rho+1 )\right]+{\cal{O}}(\rho^{3}),\ \
\end{eqnarray}
and
\begin{eqnarray}
\label{3,b}
&&\!\!\!\!\!\!\!\!\!\!\!\!\!\!
\dot{H}=\frac{1}{4(3\lambda-1)}\Big[(\omega+1)\rho(3\rho-1)
\nonumber\\
&& \ \ \ \ \ \  \ \ \   \ \ \ \,\
+8\left( \frac{k}{a^{2}}+\frac{k^2}{|\Lambda_{w}|a^{4}}\right)(3
\rho^{2}-2\rho+1)
  \Big]
\nonumber\\
&&\!
-3(\omega+1)\rho(1+\rho)H^{2}+{\cal{O}}
(\rho^{3})~,
\end{eqnarray}
where we have neglected terms of the order $\rho^{3}$ and beyond. This means that
only terms up to contribution of âdark radiationsâ, i.e $a^{-4}$, are kept. Also
it is easy to check that in the leading order approximation, these reproduce the
Friedmann
equations in standard Ho\v{r}ava-Lifshitz gravity model. Additionally, in the
above expressions we have considered the matter fluid to correspond to the standard
barotropic one, with equation of state of the form $p=\omega\rho$.

Let us now focus on the ESU. This is described by $a=a_{0}$ and $\rho=\rho_{0}$, and
hence the above Friedmann equations become
\begin{eqnarray}
\label{4,b}
&&\!\!\!\!\!\!\!\!\!\!\!\!\!\!\!\!\!
\frac{1}{6(3\lambda-1)}\left[\rho_{0}-\frac{5}{2}\rho_{0}^{2}
+3\Lambda_{w}(\rho_{0}^2 -\rho_{0}+ 1)
\right. \nonumber\\
&&
\left.
+ 6\left(\frac{k^{2}}{\Lambda_{w}a_0^{4}}-\frac{2k}{a_0^{2}}\right)
 (3 \rho_0^2-2\rho_0+1 )\right]=0,
\end{eqnarray}
\begin{eqnarray}
\label{5,b}
&&\!\!\!\!\!\!\!\!\!\!\!\!\!\! \!\!
\frac{1}{4(3\lambda-1)}\Big[(\omega+1)\rho_{0}(3\rho_0-1)
 \nonumber\\
&&\!\!
+8\left( \frac{k}{a_0^2}+\frac{k^2}{|\Lambda_{w}|a_0^{4}}\right)(3
\rho_0^2-2\rho_0+1)\Big]=0.
\end{eqnarray}

For the case of spatially closed universe, i.e. for $k=+1$, equation
(\ref{5,b}) has two roots, namely
\begin{equation}
\label{1}
\left(\frac{1}{a_{0}^{2}}\right)_{1,2}=\frac{|\Lambda_{w}|}{2}\left(-1\pm\sqrt{1-\frac{
g(\omega) }
{|\Lambda_{w}|g_0}}\right),
\end{equation}
with $g(\omega)$ and $g_0$ defined as
\begin{eqnarray}\label{2l}
&&g(\omega)\equiv (3\rho_{0}^{2}-\rho_{0})\big(\omega+1\big)\nonumber\\
&&
g_0\equiv (6\rho_{0}^
{2}-4\rho_{0}+2)~.
\end{eqnarray}
These two solutions correspond to two critical points. As we can straightforwardly see,
for any barotropic parameter $\omega$ apart from $\omega=-1$, namely apart from a simple
cosmological constant, both solutions (\ref{1}) become unphysical since they
lead to $\left(\frac{1}{a_{0}^{2}}\right)_{1,2}<0$.
Similarly, for $\omega=-1$ the two critical points  (\ref{1}) become
\begin{eqnarray}
\label{3}
\left(\frac{1}{a_{0}^{2}}\right)_{1}=0,\quad\quad
\left(\frac{1}{a_{0}^{2}}\right)_{2}=-|\Lambda_{w}|~,
\end{eqnarray}
which are both unphysical. Hence, none of these
critical points signals the presence of an ESU in closed geometry, and thus they do not
deserve further investigation.

For the case of open universe, i.e. for $k=-1$, equation
(\ref{5,b})
has
two roots, namely
\begin{eqnarray}
\label{2}
\left(\frac{1}{a_{0}^{2}}\right)_{1,2}=\frac{|\Lambda_{w}|}{2}\left(1\pm\sqrt{1-\frac{
g(\omega)}
{|\Lambda_{w}|g_0}}\right),
\end{eqnarray}
where
\begin{eqnarray}\label{2lbbb}
&&g(\omega)\equiv (3\rho_{0}^{2}-\rho_{0})\big(\omega+1\big)\nonumber\\
&&
g_0\equiv - (6\rho_{0}^
{2}-4\rho_{0}+2)~.
\end{eqnarray}
 Equation (\ref{2})
will result in two real solutions provided that
\begin{eqnarray}
\label{5}
&&
3\bigg[(\omega+1)+2|\Lambda_{w}|\bigg]\rho_{0}^{2}
\nonumber\\
&& +\bigg[9(\omega+1)+4|\Lambda_{w}|\bigg]\rho_{0}
+2|\Lambda_{w}|\geq0~.
\end{eqnarray}
For the case $\omega=-1$ the above inequality holds for any value of $\rho_0>0$
and $|\Lambda_{w}|$, and thus we obtain two critical points, namely
\begin{eqnarray}
\label{6}
\left(\frac{1}{a_{0}^{2}}\right)_{1}=0,\quad\quad
\left(\frac{1}{a_{0}^{2}}\right)_{2}=|\Lambda_{w}|~,
\end{eqnarray}
and thus only the latter is physical.
By replacing it into (\ref{4,b}) we obtain the corresponding   energy density as
\begin{equation}
\label{6*}
\!\!
(\rho_{0})_2\!=\!\frac{39|\Lambda_{w}|-1}{104|\Lambda_{w}|-5}
\left[1\pm\sqrt{1-\frac{(4788|\Lambda_{w}|-210)}{(1-39|\Lambda_{w}|)^{2}}}\right].
\end{equation}
Both of them are physically acceptable (i.e. $(\rho_{0})_2>0$) with the
condition
\begin{eqnarray}
\label{6**}
|\Lambda_{w}|>3~,
\end{eqnarray}
however, and in order to be consistent with the expansion  in terms of powers of
$\rho$ in the background dynamic equations (\ref{1,b}) and (\ref{3,b}), we keep only
the minus branch of (\ref{6*}). On the other hand, for $\omega<-1$ and $\omega>-1$, by
requiring  the physical condition $\rho_0>0$, we obtain the
following bounds on the value of $|\Lambda_{w}|$, namely
\begin{eqnarray}\label{6b}
-\frac{\omega+1}{2}<|\Lambda_{w}|<-(\omega+1)~,
\end{eqnarray}
and
\begin{eqnarray}
\label{7}
0<|\Lambda_{w}|<\frac{5(\omega+1)}{2}~,
\end{eqnarray}
respectively, along with the common lower bound on the value of $\rho_{0}$:
\begin{eqnarray}
\label{7a}
&&
\!\!\!\!\!\!\!\!\!\!\!\!\!\!\!\!\!\!\!
\rho_{0}\geq\frac{1}{6(2|\Lambda_{w}|+\omega+1)}\left\{-9(\omega+1)+4|\Lambda_{w}|
\right. \nonumber\\
&&\!\!\!\!\!\!\!\!
\left.+\sqrt{81(\omega+1)^{2}+48|\Lambda_{w}|(\omega+1)-32|\Lambda_{w}|
^{2}}\right\}\!.
\end{eqnarray}
Hence, in summary, both solutions of (\ref{2})   are physical   when the
constraints (\ref{6b})-(\ref{7a}) are fulfilled. Finally, by inserting the solutions
(\ref{2}) into equation (\ref{4,b})  we acquire
\begin{eqnarray}
\label{7aa}
\!\!\!\!\!\!\!\!\!\!\!\!\!
\left(30|\Lambda_{w}|+3\omega+\frac{1}{2}\right)\rho_{0}^{2}-\bigg(21|\Lambda_{w}
|+\omega\bigg)\rho_
0
\nonumber\\+12|\Lambda_{w}|
\pm\frac{9}{2}\sqrt{|\Lambda_{w}|^{2}g_{0}^{2}-|\Lambda_{w}|g_{0}g(\omega)}=0,
\end{eqnarray}
the solution of which will provide the corresponding $\rho_0$ for each solution. Since
analytical solutions are impossible, we solve (\ref{7aa}) numerically and in Table
\ref{Table a} we present the real and positive results for some representative values of
$\Lambda_{w}$. Note that we focus on the negative-sign branch, since only this branch is
consistent with  $-1<\omega\leq1/3$. Hence, we deduce that from the two static solutions
in (\ref{2}), only the positive-branch is compatible with  $-1<\omega\leq1/3$, namely
 \begin{eqnarray}
\label{2*}
\left(\frac{1}{a_{0}^{2}}\right)_{2}=\frac{|\Lambda_{w}|}{2}\left(1+\sqrt{1-\frac{
g(\omega)}
{|\Lambda_{w}|g_0}}\right).
\end{eqnarray}

\begin{table}
\begin{center}
 \begin{tabular}{|r|r|r|r|r|}
  \hline
&$|\Lambda_{w}|=\frac{1}{40}$&$|\Lambda_{w}|=\frac{1}{60}$&$|\Lambda_{w}|=\frac{1}{80}
$&$|\Lambda_{w}|=\frac{1}{100}$
  \\  \hline
  $\omega=-2/3$
  & $0.423$ & $0.413$ & $0.407$ & $0.404$ \\
    \hline
  $\omega=0\ \ $ &  $0.407$ & $0.403$ & $0.402$ & $0.404$ \\
    \hline
  $\omega=1/3$
  & $0.404$ & $0.403$ & $0.408$ & $0.417$ \\
  \hline
   \end{tabular}
\end{center}
\caption{\label{Tab1} The physical values of $\rho_0$ arising numerically from
(\ref{7aa}), for the negative-sign branch, for various values of $\omega$ and
$\Lambda_{w}$. }
\label{Table a}
   \end{table}

In order to study the stability of this critical point, inspired by the
first-order dynamical system approach
\cite{EllisDynamical,Copeland:1997et},
we consider the following two auxiliary variables
\begin{eqnarray}
\label{8}
y_1=a,\quad\quad y_2=\dot{a},
\end{eqnarray}
which obey the following equations:
\begin{eqnarray}\label{9}
&&\!\!\!\!\!\!\!\!\!\!\!\!\!\!\!
\dot{y_1}=y_2=f_1(y_1,y_2),\\
&&\!\!\!\!\!\!\!\!\!\!\!\!\!\!\!
\dot{y_2}=\frac{1}{4(3\lambda-1)}\Big\{
(\omega+1)\rho(3\rho-1)
 y_1\nonumber\\
&&\ \ \ \ \ \  \ \ \
+8\left(3\rho^{2}-2\rho+1\right)\Big(\frac{k}{y_1}
+\frac{k^2}{y_{1}^{3}|\Lambda_
{ w } | }
\Big)
 \Big\}\nonumber\\
&&\!\!\!
+
\Big[1-3(\omega+1)\rho(\rho+1)\Big]\frac{y_{2}
^{2}}{y_{1}}=f_2(y_1,y_2).
\end{eqnarray}
As usual, by deriving the eigenvalues square $\vartheta^2$ of the Jacobian matrix
\begin{eqnarray}\label{10}
J\bigg(f_1(y_1,y_2),f_2(y_1,y_2)\bigg)=\left(
\begin{array}{cc}
\frac{\partial f_1}{\partial y_1} & \frac{\partial f_1}{\partial y_2}  \\
\frac{\partial f_2}{\partial y_1} & \frac{\partial f_2}{\partial y_2}  \\
\end{array}
\right),
\end{eqnarray}
one can deduce on the stability of the critical points.
If $\vartheta^2<0$ then the corresponding critical point
can be interpreted as a stable center point. Since, as it was mentioned above, for the
closed universe there is not a physically accepted static solution, in the following we
focus on the open case.

\begin{figure}[ht]
\begin{tabular}{c}
\epsfig{figure=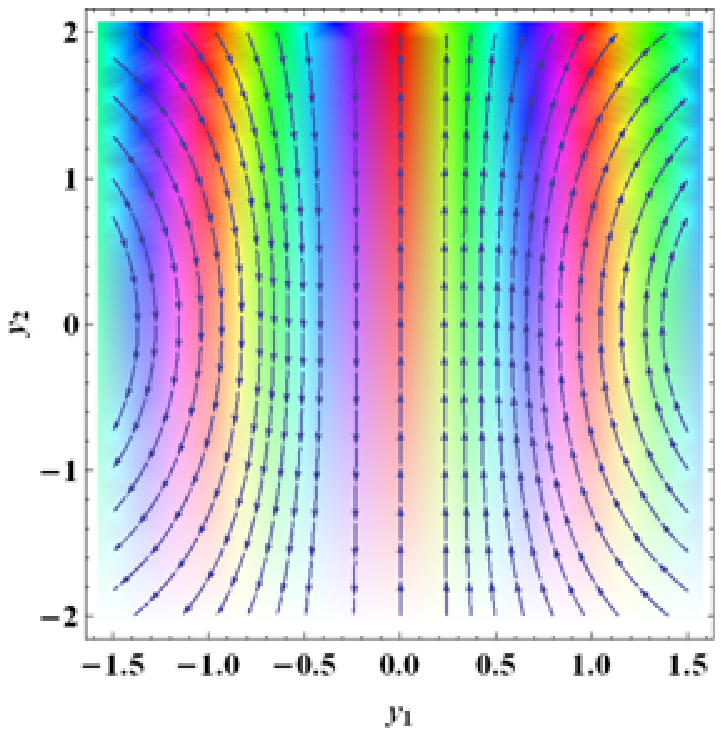,width=5.5cm}\\
\epsfig{figure=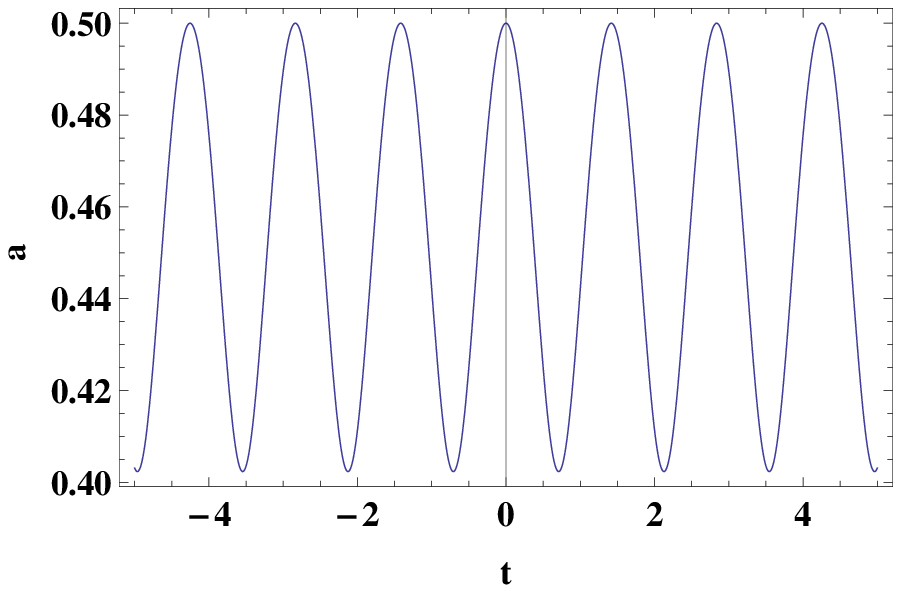,width=6cm}
\end{tabular}
\caption{{\it{ The phase diagram in  $(a,\dot{a})$ or $(y_1,y_2)$
space (upper graph) and the evolution of the scale factor in terms
of time (lower graph) for the spatially open cosmology, with equation-of-state
parameter $\omega=-1$. We have set $\lambda=1$ and $ |\Lambda_{w}|=5 $, while for
$\rho_{0}$ we have used the values obtained numerically from the negative-sign branch of
  (\ref{6*}), namely $\rho_0=0.148$.  }}}
\label{fig0a}
\end{figure}

In the case of $\omega=-1$, after a simple
calculation  we acquire:
\begin{eqnarray}
 &
&\!\!\!\!\!\!\!\!\! \!
\mbox{Critical point }
\left(\frac{1}{a_{0}^{2}}\right)_{2}=|\Lambda_{w}|,\nonumber\\
&&\!\!\!\!\!\!\!\!\! \!
\Longrightarrow\
 \vartheta^2= \big(-12\rho_{0}^{2}+8\rho_0-4\big)\frac{|\Lambda_{w}|}
 {(3\lambda-1)},
\end{eqnarray}
and thus it implies the stability of ESU  corresponding
to the physical critical point $(y_1,0)$, since
$\vartheta^2|_{(\rho_{0})_{2}}<0$. In order to see this
feature more transparently, in the upper graph of Fig. \ref{fig0a}
we depict the phase-space
behavior for the open geometry  for
$\omega=-1$.    Additionally, in order to verify the stability of the ESU in an
alternative way, we
add a small deviation to the scale-factor value
$(\frac{1}{a_{0}^{2}})_2=|\Lambda_{w}|$ and in    the lower graph of Fig.
\ref{fig0a} we depict its evolution, as it arises numerically from (\ref{9}). As we
observe the universe exhibits small oscillations around the scale-factor value of the
ESU, without deviating from it, as expected.

In the case of $\omega\neq-1$, we acquire:
\begin{eqnarray}
\label{10**}
 \vartheta^2=\frac{1}
 {4(3\lambda-1)}\left [(\omega+1)(3\rho_{0}^{2}-\rho_{0})
 \right. \nonumber\\ \left.+8\bigg(3\rho_{0}
 ^{2}-2\rho_0+2\bigg)
 \left(\frac{1}{a_{0}^{2}}-\frac{3}{|\Lambda_{w}|a_{0}^{4}}\right)\right],
\end{eqnarray}
with $a_{0}$ given in  (\ref{2*}) and  $\rho_{0}$ arising from
(\ref{7aa}), i.e from Table \ref{Table a}.
In Table \ref{Table b} we provide the 
corresponding values of $\vartheta^2$. As we observe,  $\vartheta^2<0$ for all values of 
$|\Lambda_{w}|$,$\omega$ and 
($a_0,\rho_0$) consistent with $0\leq\rho_0<1$ and  $-1<\omega\leq1/3$, and thus we 
deduce that  the ESU 
is stable. 
\begin{table}[!]
\begin{center}
 \begin{tabular}{|r|r|r|r|r|}
  \hline
  &$|\Lambda_{w}|=\frac{1}{40}$&$|\Lambda_{w}|=\frac{1}{60}$&$|\Lambda_{w}|=\frac{1}{80}$
   &$|\Lambda_{w}|=\frac{1}{100}$
 \\  \hline
 $\omega=-\frac{2}{3}$ & $\vartheta^2=-0.37$ & $\vartheta^2=-0.32$ & $\vartheta^2=-0.3$ &
$\vartheta^2-0.28$ \\
    \hline
 $\omega=0$ & $\vartheta^2=-0.86$ & $\vartheta^2=-0.82$ & $\vartheta^2=-0.78$ & 
$\vartheta^2=-0.76$ 
\\
   \hline
 $\omega=\frac{1}{3}$ & $\vartheta^2=-1.1$ & $\vartheta^2=-1.06$ & $\vartheta^2=-1.01$ &
$\vartheta^2=-0.98$\\
  \hline
  \end{tabular}
\end{center}
\caption{\label{Tab2}
The eigenvalues corresponding to the ESU critical points
($a_0,\rho_0$) obtained from (\ref{2*}) and the values of $\rho_{0}$ given in Table
\ref{Table a}, for
$\lambda=1$. }
\label{Table b}
 \end{table}
However, although a stable ESU is easily realized, in order to obtain a full realization
of the emergent universe scenario we need an additional mechanism that could make the
universe deviate from ESU after a large time interval, and enter
into the usual expanding thermal history. This would be possible only in the presence of
an exotic matter sector, with equation-of-state parameter outside the range
$-1<\omega\leq1/3$.

\section{Scalar perturbations and stability conditions}
\label{perturbationsanalysis}

In this section we perform an analysis of the scalar perturbations in
Ho\v{r}ava-Lifshitz-like $F(R)$
gravity
in a cosmological framework. In particular, we desire to extract conditions under which
the scenario at hand is free of perturbative instabilities, and hence
physical\footnote{Note that stability in this section is used in a different sense than
that of dynamical system framework, as in the previous section. In particular, in
dynamical system analysis an unstable solution is one that cannot attract the universe,
however it is completely physical. On the other hand, in perturbation
analysis, a solution with perturbative instabilities implies that it is ill-behaved and
not physical.}. In order to achieve this, we
linearly perturb equations (\ref{1,b}) and (\ref{3,b}) around the static states
(\ref{4,b}) and (\ref{5,b}). Applying the following perturbations in the scale factor
and matter density:
\begin{eqnarray}
\label{6,b}
&&a(t)\rightarrow a_{0}(1+\delta a(t)),\nonumber\\
&&\rho(t)\rightarrow \rho_{0}(1+\delta \rho(t))~,
\end{eqnarray} linearizing using
\begin{eqnarray}
\label{8,b}
&&(1+\delta a(t))^{n}\simeq 1+n\delta a(t),\nonumber\\
&&(1+\delta \rho(t))^{n}\simeq 1+n\delta \rho(t),
\end{eqnarray}
and imposing the background Friedmann equation (\ref{4,b}) in order to eliminate
background quantities, we obtain
\begin{eqnarray}
\label{10,b}
&&\!\!\!\!\!\!\!\!\!\!\!\!\!
\rho_{0}\delta \rho(t)\Big\{
1-5\rho_{0}+6\Lambda_{w}(\rho_{0}-1) \nonumber\\
&& \ \ \ \ \
+12\Big(\frac{2k}{a_{0}^2}-\frac{k^{2}}{\Lambda_{w}a_{0}^{4}}\Big)
(1-3 \rho_{0})
 \Big\}=
\nonumber\\
&&\!\!\!\!\!\!\!\!\!\!\!\!\!
-\delta
a(t)\Big\{24\Big(\frac{k}{a_{0}^{2}}-\frac{k^{2}}{\Lambda_{w}a_{0}^{4}}
\Big)(1-2\rho_{0}+3 \rho_{0}^{2})
 \Big\}.
\end{eqnarray}
Similarly, perturbing equation (\ref{3,b}), linearizing, and imposing (\ref{4,b}) in
order to eliminate background quantities, we acquire
\begin{eqnarray}
\label{13,b}
&&\!\!\!\!\!\!\!\!\!\!
\delta\ddot{a}=\frac{1}{4(3\lambda-1)}\Big\{ \rho_{0}\delta\rho
\Big[(\omega+1) (2\rho_{0}-1)
\nonumber\\
&&\ \ \
+\frac{8k}{a_{0}^{2}}\rho_{0}(1+\omega)
-16\Big(\frac{k}{a_{0}^{2}}+\frac{k^{2}}{
|\Lambda_{w}
|a_{0}^{4}}
\Big)(1-3 \rho_{0})\Big]  \nonumber\\ &
&\ -
\delta
a\Big[
16\Big(\frac{k}{a_{0}^{2}}+\frac{2k^{2}}{|\Lambda_{w}|a_{0}^{2}}
\Big)(1-2 \rho_{0}+3 \rho_{0}^{2})
 \Big]\Big\}\;.
\end{eqnarray}

Thus, using (\ref{10,b}) in order to find $\rho_{0}\delta\rho$ in terms of
$\delta a$, and substituting into (\ref{13,b}), leads to the following
differential equation:
\begin{eqnarray}
\label{14,b}
\delta\ddot{a}-\frac{\left(A\times B\times C-D\right)}{(3\lambda-1)}\delta a=0\;,
\end{eqnarray}
where we have defined
\begin{eqnarray}
\label{15,b}
&&\!\!\!\!\!\!\!\!\!\!\!\!
A\equiv
-(6\rho_{0}\Lambda_{\omega}+\rho_{0}+1)
+\frac{24k}{a_{0}^{2}}(1+3\rho_{0}^{2})
 \nonumber\\
 &&
-\frac{12k^{2}}{\Lambda_{w}a_{0}^{4}}(1-3 \rho_{0})+
36\left(\Lambda_{\omega}-\frac{4k}{a_{0}^{2}}+\frac{2k^{2}}{\Lambda_{w}a_{0}^{4}}
\right)^{2}
 \nonumber\\
 &&
 +\left(5-6\Lambda_{\omega}+\frac{72k}{a_{0}^{2}}-\frac{
36k^{2}}{a_{0}^{2}}
\right)\rho_{0}
\nonumber\\
&&\ \ \
\cdot
\left(\rho_0-12\Lambda_{\omega}+\frac{48k}{a_{0}^{2}}-\frac{24k^{
2}}{\Lambda_{w}a_{0
}^{4}}\right),
\end{eqnarray}
 \begin{equation}
 \label{16,b}
B\equiv
 2\left(\frac{k}{a_{0}^{2}}-\frac{k^{2}}{|\Lambda_{w}|a_{0}^{4}}\right)(1-2\rho_{0}+3
\rho_{0}^{2}),
\end{equation}
\begin{equation}
\label{17,b}
 C\equiv
3(1+\omega)(2\rho_{0}-1)-48\left(\frac{k}{a_{0}^{2}}+\frac{k^{2}}{|\Lambda_{w}|a_
{ 0}^{4}}\right) (1-3\rho_{0}),
\end{equation}
and
\begin{eqnarray}
\label{18,b}
D&\equiv&
-4\left(\frac{k}{a_{0}^{2}}+\frac{2k^{2}}{|\Lambda_{w}|a_{0}^{4}}\right)
(1-2\rho_{0}+3
\rho_{0}^{2}).
 \end{eqnarray}
Hence, from the differential equation (\ref{14,b}) we deduce that in order to have a
stable ESU against scalar perturbations in the framework of Ho\v{r}ava-Lifshitz-like
$F(R)$ gravity, the
following condition must be satisfied:
\begin{eqnarray}
\label{19,b}
\frac{\left(A\times B\times C-D\right)}{3\lambda-1}<0\;.
\end{eqnarray}
 
As usual, the instabilities-absence condition (\ref{19,b}) must be applied in the 
background solutions of the model, extracted in the previous section. Since we are 
dealing with Einstein static universe, these solutions were characterized only by 
$a_{0}, \rho_{0}, k$, with  $\omega$ and $\Lambda_{w}$ the model parameters, and were 
summarized in expressions (\ref{6}), (\ref{6*}), (\ref{2*}) and Table \ref{Table a}.
Since in the case of a closed universe ESU was not realized, we focus on the case of open 
geometry. For the case $\omega=-1$, by setting the background configuration 
$(a_{0},\rho_{0})$ from equations (\ref{6}) and (\ref{6*}), subject to the constraint 
(\ref{6**}), and substituting them into equations (\ref{15,b})-(\ref{18,b}),
we immediately find that inequality (\ref{19,b}) holds. Similarly, in the case  
$\omega\neq -1$, substituting the values of $a_0$ and $\rho_0$ from (\ref{2*}) and Table I 
into (\ref{15,b})-(\ref{18,b}) we deduce that inequality (\ref{19,b}) is satisfied 
too. Hence, in summary, we can see that the ESU obtained in 
the precious section is free of perturbative instabilities.

\section{Remarks and Conclusions}
\label{Conclusions}

In this work we performed an investigation of the Einstein static universe (ESU) in the
framework of Ho\v{r}ava-Lifshitz-like $F(R)$ gravity. Such a gravitational modification
is obtained by employing higher-order $R$-terms, keeping both the detailed balance and
projectability conditions, and although contrary to the  usual $F(R)$ gravity  it
is not full $D + 1$ diffeomorphism invariant, in the limit of linear
approximation it recovers the usual Ho\v{r}ava-Lifshitz counterpart. Hence,
 we were interested in examining whether the cosmological application of this
theory allows for the realization of ESU, which is the basic concept in the realization
of the emergent universe scenario. If this is the case, then the
initial-singularity problem of standard Big-Bag cosmology can be alleviated.

As a first step we performed a dynamical analysis of Ho\v{r}ava-Lifshitz-like $F(R)$
cosmology in the phase space. We showed that in the case of closed geometry there is no
stable  physically meaningful ESU, while in the case of open geometry the ESU can be an
attractor in the presence of both exotic and usual matter. However, the most physically
interesting result was that in the case of open geometry ESU is stable and thus it can be
realized. Nevertheless, in order to obtain a full realization
of the emergent universe scenario we need an additional mechanism that could make the
universe deviate from ESU after a large time interval, and enter
into the usual expanding thermal history, which can be obtained only  through an exotic
matter sector with unconventional equation-of-state parameter.

As a second step we examined the behavior of ESU under scalar perturbations, desiring to 
extract the conditions under which the scenario at hand is free of perturbative 
instabilities such as ghosts or Laplacian instabilities. Our analysis showed that the 
background ESU solutions are free of perturbative 
instabilities.

The above results imply that ESU can be safely realized in the framework of
Ho\v{r}ava-Lifshitz-like $F(R)$ cosmology, however the emergent universe scenario is not
straightforward in such a gravitational modification, since an exotic form of matter is
required. Hence,  within the same theory we have both a cosmological
advantage, namely that we alleviate the initial-singularity problem, as well as a
theoretical advantage, namely that the underlying theory has an improved
renormalizable behavior in the UV. These features make the above construction a good
candidate for the description of Nature, that is worthy to be investigated further.

\section*{Acknowledgments}

We would like to thank the anonymous referee whose careful and useful comments led to a very improved revision of the paper. This work has been supported financially by Research Institute for Astronomy and
Astrophysics of Maragha (RIAAM) under research project No.1/4165-93.

\end{document}